\documentclass[%
 reprint,
 amsmath,amssymb,
 aps,
]{revtex4-1}
\usepackage{float}
\usepackage[colorlinks=true,linkcolor=blue,citecolor=blue]{hyperref}%
\usepackage{xcolor}
\usepackage{graphicx}
\usepackage{dcolumn}
\usepackage{bm}
\usepackage{fancyhdr}
\begin{document}


\title{Role of net baryon density on rapidity width of identified particles from the lowest energies available at the CERN Super Proton Synchrotron to those at the BNL Relativistic Heavy Ion Collider and at the CERN Large Hadron Collider}
\author{Nur Hussain}
 \email{nur.hussain@cern.ch}
\author{Buddhadeb Bhattacharjee}%
 \email{Corresponding author; buddhadeb.bhattacharjee@cern.ch}
\affiliation{%
Nuclear and Radiation Physics Research Laboratory, Department of Physics, Gauhati University, Guwahati 781014, India
}%

\date{\today}

\begin{abstract}
Widths of the rapidity distributions of various identified hadrons generated with the UrQMD-3.4 event generator at all the Super Proton Synchrotron (SPS) energies have been presented and compared with the existing experimental results. An increase in the width of the rapidity distribution of $\Lambda$ could be seen with both Monte Carlo (MC) and experimental data for the studied energies. Using MC data, the study has been extended to Relativistic Heavy Ion Collider (RHIC) and Large Hadron Collider (LHC) energies. A similar jump, as observed in the plot of rapidity width versus rest mass at Alternating Gradient Synchrotron (AGS) and all SPS energies, persists even at RHIC and LHC energies, confirming its universal nature from AGS to the highest LHC energies. Such observation indicates that pair production may not be the only mechanism of particle production at the highest LHC energies. However, with MC data, the separate mass scaling for mesons and baryons is found to exist even at the top LHC energy.

\end{abstract}

\maketitle
\bibliographystyle{apsrev4-1} 
\bibliography{xampl}


\section{\label{sec:level1}INTRODUCTION}
Widths of the rapidity distributions of particles are considered to be a measure of their final-state rescattering and are found to follow a mass ordering [\onlinecite{Blume},\onlinecite{Aichelin}]. In addition to the final-state rescattering, the rapidity width is also sensitive to a number of other parameters such as longitudinal dynamics, the velocity of sound, etc. [\onlinecite{Klay},\onlinecite{Bedanga}]. Recently, in Ref.\ [\onlinecite{Deysir}], it has been shown that the width of the rapidity distribution of a particle, in addition to its mass dependence, also depends on the nature of its quark content. In that report, from a study on ultrarelativistic quantum molecular dynamics (UrQMD)  generated Au+Au data at 10, 20, 30, and 40$\it{A}$ GeV and its comparison with NA49 experimental results, it has been shown that the width of the rapidity distribution of a particle like $\Lambda$ ($uds$), containing leading quarks, exhibits an additional dependence on net baryon density distribution, whereas particles not containing leading quarks, such as $\bar{\Lambda}$ ($\bar{u} \bar{d} \bar{s}$), do not exhibit any such dependence on net baryon density distribution. In rapidity space, the $\Lambda$ distribution is shown to reflect the pattern of net baryon density distribution, whereas $\bar{\Lambda}$ does not.  Since the net baryon density distribution depends on the transparency of collision, evolution of rapidity width of identified particles with energies from the lowest Super Proton Synchrotron (SPS) energy to the highest Relativistic Heavy Ion Collider (RHIC) and Large Hadron collider (LHC) energies is expected to provide some important information about the role of net baryon density distribution on rapidity width at different energy densities and temperatures and as such on the particle production mechanism itself. In this paper, an attempt has therefore been made to trace this signature of net baryon density distribution on rapidity width over the entire SPS energies such as 20, 30, 40, 80, and 158$\it{A}$ GeV with UrQMD-3.4-generated and available experimental data and to extend the study with Monte Carlo (MC) data to RHIC and LHC energies. The ultimate motivation of the present investigation is to check if this dependence of rapidity width on net baryon density distribution is a general characteristic of the rapidity distribution of the produced particles of heavy ion collisions.
\begin{figure*}[htbp]
		\centering
		\includegraphics[width=7in]{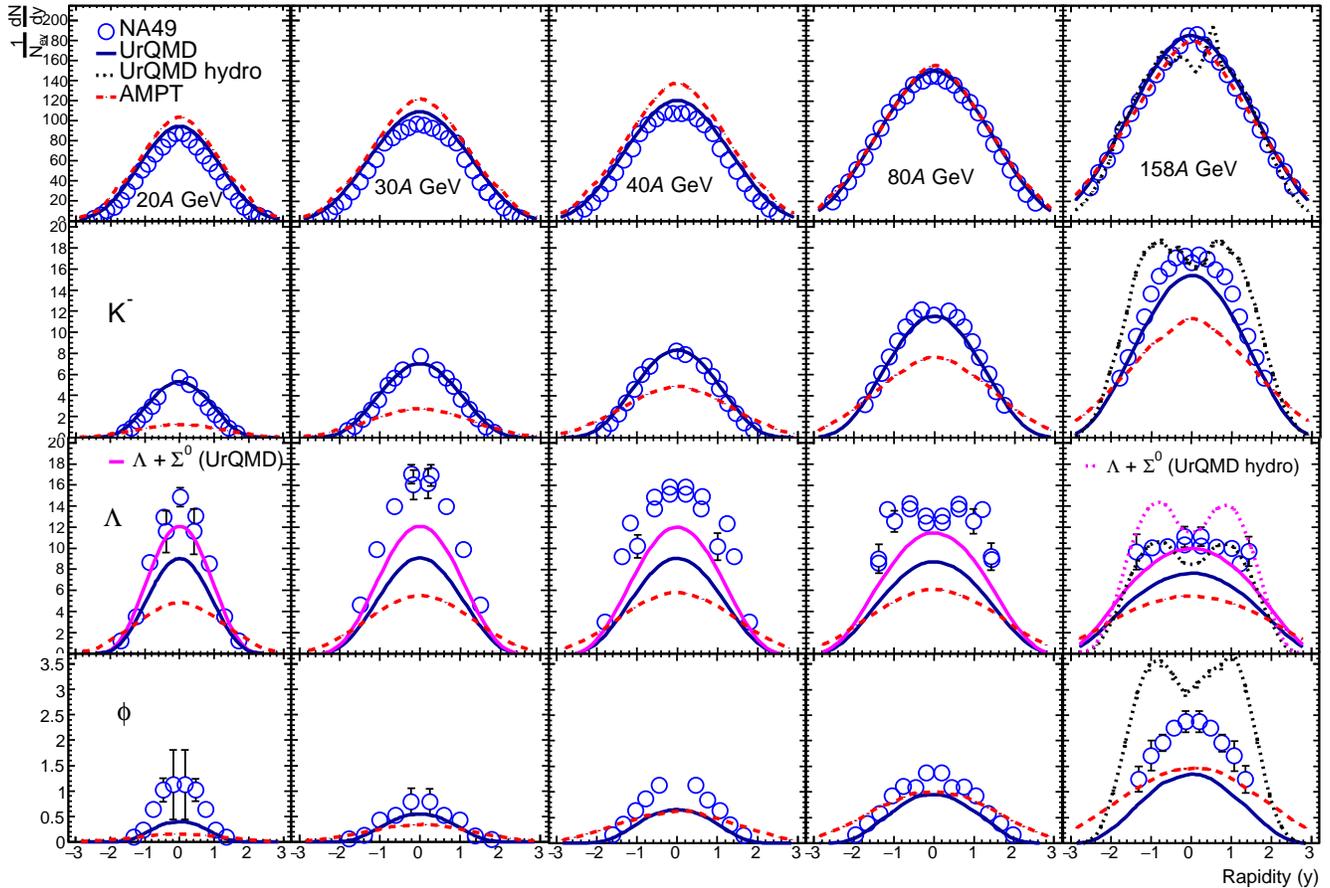}
		\caption{(Color online) Comparison of experimental (NA49) rapidity distributions of $\pi^-,  K^-, \phi$, and $ \Lambda $ in Pb-Pb collisions with UrQMD-3.4, AMPT (string melting) generated data at $E_{lab}$ $= 20, 30, 40, 80$, and $158\it{A}$ GeV  and UrQMD (Hydro) at $158\it{A}$ GeV [\onlinecite{Blume},\onlinecite{Afanasiev}--\onlinecite{Afanasiev2}]. The blue open circular symbols represent the experimental results of NA49, whereas the blue solid, black dotted, and red dashed curves denote the results of UrQMD-3.4, UrQMD-3.4 (Hydro), and AMPT (string melting) data respectively. The magenta solid and dotted curves represent the rapidity distributions of $ \Lambda + \Sigma^0$ using UrQMD-3.4 and UrQMD-3.4 (Hydro) data.}
		\label{Fig:RapiditySPS}
\end{figure*}
\section{\label{sec:level2}RESULTS}
The ultrarelativistic quantum molecular dynamics (UrQMD) event generator is a many-body microscopic Monte Carlo model for generating events of $pp$, $p-A$ and $A-A$ collisions at relativistic and ultrarelativistic energies. The recent versions of UrQMD models include \textsc{pythia} to incorporate hard perturbative QCD effects [\onlinecite{Bleicher},\onlinecite{Bass}]. The UrQMD model is based on mesonic and baryonic degrees of freedom associated with constituent quarks, diquarks, and covariant propagation of color strings [\onlinecite{Mitrovski}]. It also includes the excitation and fragmentation of color strings, formation and decay of hadronic resonances, and rescattering of particles. UrQMD could be applied successfully via fragmentation of strings and hard scattering to explore heavy ion reactions from AGS to SPS energies  ($E_{lab}$ $= 20\it{A} $ to $158\it{A}$ GeV) and the full RHIC energies ($\sqrt{\it{s}_{NN}}$ = $19.6 $ to $200$ GeV) [\onlinecite{Bravina},\onlinecite{Bravina2}] . In Ref.\ [\onlinecite{Mitrovski}], from a comparison on UrQMD calculations with experimental results of $p+\bar{p}$, Pb + Pb, and Au + Au collisions from 17.3 GeV at the SPS to 1.8 TeV at Fermilab, it has been shown that UrQMD has a valid basis for further extrapolation in energy to make predictions at LHC energies. Therefore, to start with, events were generated using the latest version, UrQMD-3.4, for Pb-Pb collisions at all the beam energies of SPS at $E_{lab}$ $= 20, 30, 40, 80$, and $158\it{A}$ GeV. The event statistics of generated data for minimum biased events is shown in the Table~\ref{tablesps1}.

\begin{center}
\setlength{\tabcolsep}{1em}
\begin{table*}[htbp]
\caption {Event statistics of UrQMD-3.4-generated data for various SPS energies.} \label{tablesps1}
\centering
\begin{tabular}{ c c c c c c c c c}
\hline
\hline
 System & Energy &  Events & $\pi^- \times 10^8  $  & $ K^- \times 10^6  $ &  $\phi \times 10^5 $ & $\Lambda \times 10^7  $ & $\Xi^- \times 10^5  $ & $\Omega^- \times 10^3 $    \\ 
 &$( \it{A} $ GeV)&  (million) &  &  &  &   &&  \\
 \hline
 Pb - Pb &	     20       &     1.49       &     1.040    &     4.05     &    2.173  &     6.94     &    2.74     &   3.60 \\
                     & 30    &      1.25       &     1.06      &    4.92    &       2.81    &     6.60     &   2.85    &   5.07   \\
                     &40      &      1.30      &     1.27       &    6.46    &     3.79     &   7.42       &   3.40     &   7.21   \\
                     &80      &       0.40     &    0.59        &   3.55      &    2.26      &    2.98      &  1.56      &  4.86  \\
                     &158    &        0.76    &      1.27      &     8.61     &    5.86       &  5.62        & 3.32       & 13.89 \\
 \hline
 \hline
\end{tabular}
\end{table*}
\end{center}
\begin{center}
\setlength{\tabcolsep}{1em}
\begin{table*}[htbp]
\caption {Widths of the rapidity distributions of all the studied hadrons as estimated using Eq. (\ref{eq:1}) for UrQMD-3.4-model-generated data and experimental data. While the experimental values of widths (r.m.s.) of $\pi^-,  K^-, \phi$, and $\Omega^-$ are taken from Ref.\ [\onlinecite{Blume}] and  $\Lambda, \bar{\Lambda}$, and $ \Xi^- $   are taken from Ref.\ [\onlinecite{Alt}].} 
\label{tableSPSWidth2}
\centering
\begin{tabular}{ c c c c c c}
\hline
\hline    
 Particles                                  & $20\it{A} $ GeV           &  $30\it{A} $ GeV             & $40\it{A} $ GeV            &  $80\it{A} $ GeV            & $158\it{A} $ GeV   \\ 
 \hline 
 $\pi^-$                                     & $1.063\pm0.00098$   &    $ 1.133\pm 0.0012$     &  $ 1.185\pm0.0014$           & $ 1.317\pm0.0047  $    & $ 1.441\pm0.0065$  \\
                                                 &($0.991\pm0.01)(data)$    & ($ 1.068\pm0.01 $)           &  ($ 1.123\pm0.01 $)      & $ 1.288\pm0.01     $    & ($ 1.432\pm 0.01   $)  \\
 $K^-$                                     & $0.816\pm0.0027$            &    $ 0.890\pm0.0038$      &  $ 0.944\pm0.0045$       & $ 1.062\pm0.0067$      & $ 1.167\pm0.0036$  \\
    					       &($0.727\pm0.034 $)          &    ($ 0.798\pm0.009 $)      &  ($ 0.852\pm0.069 $)      & ($ 0.9711\pm0.01   $)    & ($ 1.087\pm0.01 $)  \\                                           
 $\phi$			                & $0.670\pm0.0044 $           &    $ 0.751\pm0.0051$     &  $ 0.805\pm0.0049$        & $ 0.937\pm0.0079 $    & $1.062 \pm0.0065$  \\
 						 &($0.565\pm0.040 $)         &    ($ 0.710\pm0.095  $)     &  ($ 0.815\pm0.0701  $)       & ($ 0.851\pm0.127    $)    & ($ 1.190\pm0.176   $)  \\
 $\Lambda$				 & $0.806\pm0.0024$     &    $ 0.907\pm0.0039$         &  $ 0.977\pm0.0047$        & $ 1.133\pm0.0049 $     & $ 1.327\pm0.0065$  \\
 						 &($0.70\pm0.01 $)            &    ($0.89\pm 0.02  $)          &  ($ 1.11\pm0.08 $)           & ($ 1.28\pm0.02     $)      & ($ 1.97\pm 0.35  $)  \\
 $\Lambda + \Sigma^0$             & $0.810\pm 0.02020$   &    $0.910\pm0.0032$       &  $0.984\pm0.0040$       & $ 1.143\pm0.0042  $        & $ 1.343\pm0.0029$  \\
 
 $\bar{\Lambda}$			& $0.722\pm0.0202$       &    $0.824\pm0.038$          &  $0.935\pm0.0417$        & $ 1.108\pm0.0023  $    & $ 1.083\pm0.0854$  \\
 						 &($0.62\pm0.14 $)            &    ($0.69\pm0.05  $)        &  ($ 0.77\pm0.05$)             & ($ 0.83\pm0.05  $)      & ($ 1.00\pm0.03 $)  \\
 $\Xi^-$					& $0.671\pm 0.0050$       &    $0.754\pm 0.0075$      &  $0.808\pm0.0085$       & $ 0.946\pm0.012$       & $ 1.073\pm0.0094$  \\
 						 &($0.64\pm0.08  $)            &    ($0.73\pm0.08   $)          &  ($ 0.94\pm0.13 $)           & ($ 0.98\pm0.25   $)        & ($ 1.18\pm0.18   $)  \\
 $\Omega^-$				& $0.577\pm0.0380$       &    $0.637\pm0.024$         &  $0.664\pm0.032$         & $ 0.800\pm0.033  $    & $ 0.919\pm0.0550$  \\
 						&				    &					        & ($0.596\pm 0.09 $)     	     &	             			&($1.200\pm0.396 $)\\
  \hline
 \hline
\end{tabular}
\end{table*}
\end{center}
\begin{figure*}[htbp]
\begin{tabular}{ll}
\includegraphics[scale=0.44]{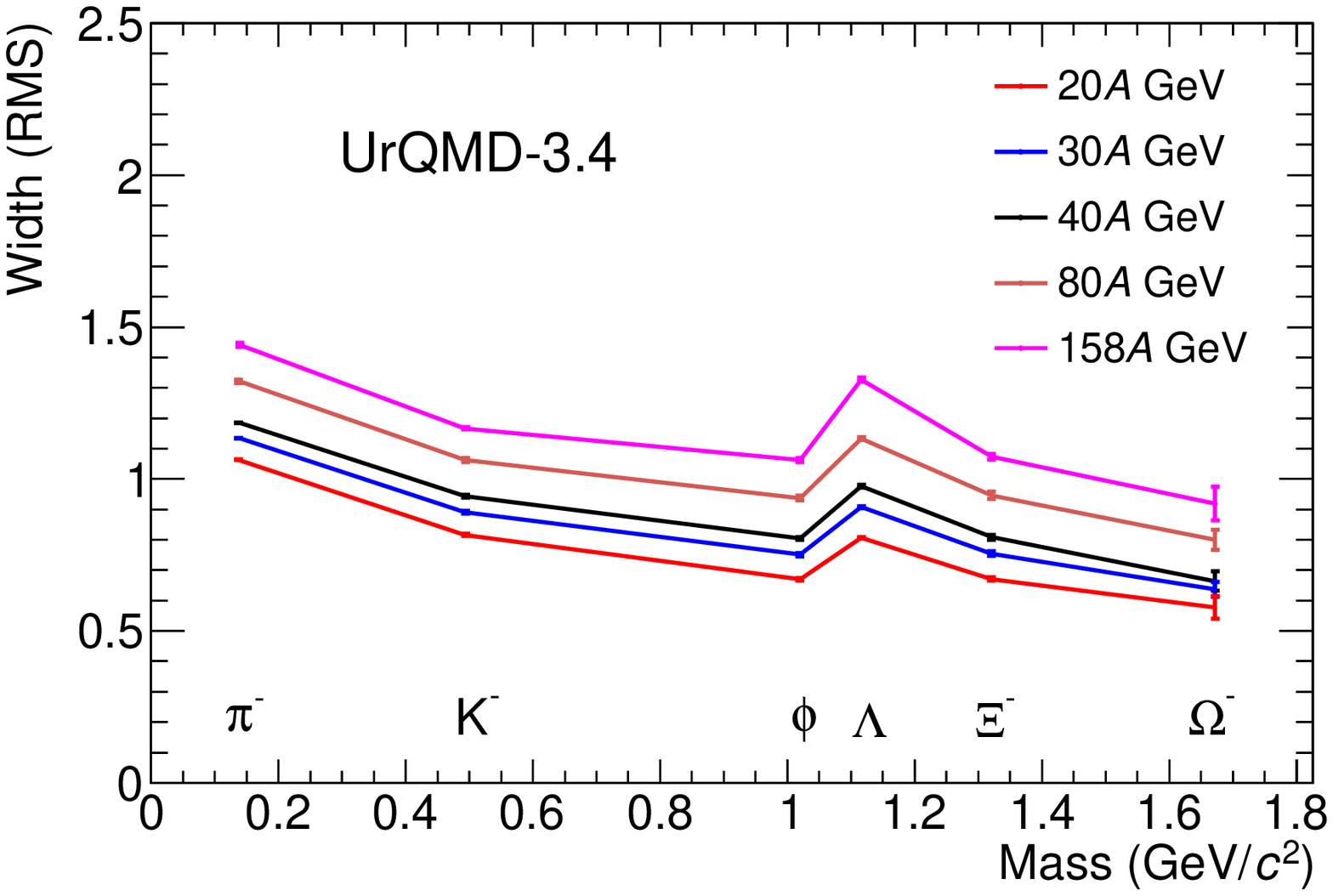}
&
\includegraphics[scale=0.44]{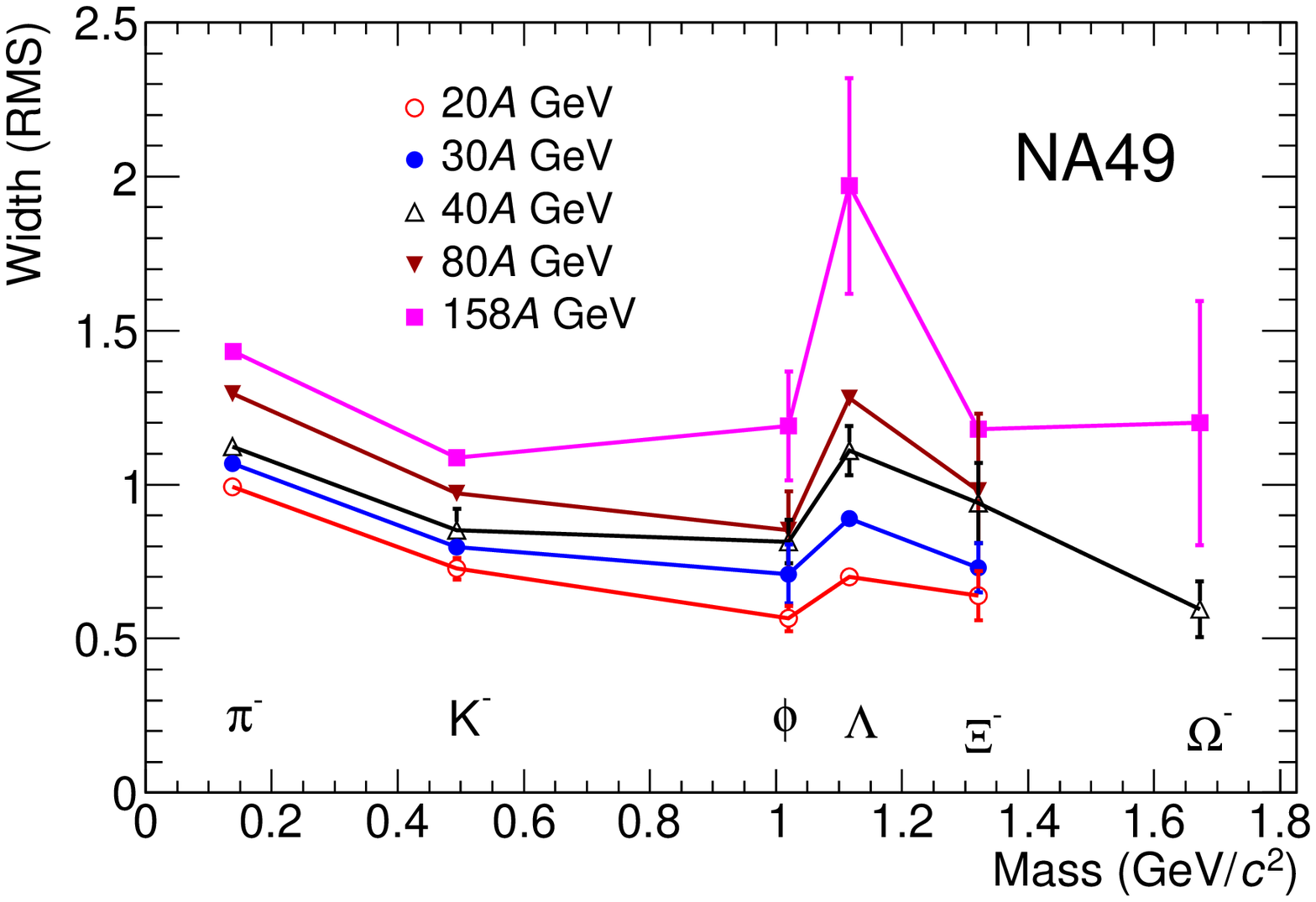}
\end{tabular}
\caption{(Color online) Width of the rapidity distribution as a function of masses of $\pi^-,  K^-, \phi, \Lambda, \Xi^-$, and $\Omega^-$ in Pb-Pb collisions at all the SPS energies using UrQMD-3.4-generated events (left panel) and NA49 results (right panel) [\onlinecite{Blume}, \onlinecite{Alt}].}
\label{Fig:widthsps}
\end{figure*}
Rapidity distributions of $\pi^-,  K^-, \phi$, and $ \Lambda $ with generated data of central Pb-Pb collisions at all the SPS energies from $E_{lab}$ $= 20\it{A} $ to $158\it{A}$ GeV are shown in Fig.~\ref{Fig:RapiditySPS} and compared with the NA49 results [\onlinecite{Blume},\onlinecite{Afanasiev}--\onlinecite{Afanasiev2} ]. While the rapidity distribution of $\pi^-$ meson of UrQMD-generated events are found to be in good agreement with experimental results of all the studied energies, the model-generated rapidity distributions of $K^-$ are found to agree well with experimental results only up to $80\it{A}$ GeV. The model somewhat underestimates the experimental values around midrapidity at the highest SPS energy, i.e.,	 $158\it{A}$ GeV. Since the experimental $\Lambda$ has a contribution from the decay of $\Sigma^0 $, UrQMD-generated $\Lambda +\Sigma^0 $ is compared with the experimental rapidity distribution of $\Lambda$. At all the SPS energies, the MC values of $\Lambda +\Sigma^0 $ are found to be less than the experimental $\Lambda$ values. However, the disagreement decreases with the increase of energy. In the case of the $\phi$ meson, UrQMD underestimates the experimental values significantly at all energies and the disagreement is more at higher energies. 

The experimental rapidity spectra of various identified particles have also been compared with the prediction of other event generators such as AMPT (string melting) at all the SPS energies and UrQMD (Hydro) at 158$\it{A}$ GeV (as the disagreement in the rapidity spectra of $\phi$  is maximum at this energy for UrQMD-generated and experimental data). It is readily evident from Fig.~\ref{Fig:RapiditySPS} that the disagreement with the experimental values is even more for data generated by both these models and hence they have not been considered for further analysis.

Rapidity distributions of UrQMD-generated and experimental  $\pi^-,  K^-, \phi$, and $ \Lambda $ at all SPS energies are parameterized by the following double Gaussian function [\onlinecite{Alt},\onlinecite{Deysir}]: 
\begin{equation} \label{eq:1}
\begin{aligned} 
\frac{dN}{dy} \propto (e^{- \frac{(y-\bar{y})^2}{2\sigma^2}} + e^{- \frac{(y+\bar{y})^2}{2\sigma^2}}),\\
\end{aligned}
\end{equation}
where the symbols have their usual significance. Widths of the rapidity distributions of all the studied hadrons, both generated and experimental, are listed in Table~\ref{tableSPSWidth2} and plotted as a function of their rest mass in Fig.~\ref{Fig:widthsps}.

It could be readily seen from Fig. \ref{Fig:widthsps} that, as reported in Ref.\ [\onlinecite{Deysir}] for AGS and low SPS energies, a jump in the widths of $\Lambda$ does exist even at higher SPS energies ($E_{lab}\ge 80\it{A} $ GeV)  for both generated and experimental data.
\begin{center}
\setlength{\tabcolsep}{1em}
\begin{table*}[htbp] 
\caption {Widths of the rapidity distributions of all the studied hadrons with UrQMD-3.4-generated events at RHIC and LHC energies.}
 \label{tableRHICLHCWidth}
\centering
\begin{tabular}{ c c c c c c c}
\hline
\hline    
 Particles                                  & $19.6$ GeV                             &  $62.4 $ GeV                         & $130$ GeV                             &  $200 $ GeV                 & $2.76 $ TeV                   & $5.02 $ TeV    \\ 
 \hline  
 $\pi^-$                                    &$1.431\pm0.00297$	              &$1.855\pm0.0001$                 &$2.127\pm0.0036	$            &$2.349\pm0.0022$        & $3.603\pm0.0010$	&$4.026\pm0.0072$\\
 $K^-$                                     & $1.191\pm0.0038$	               &$1.637\pm0.0050$	             &$1.971\pm0.0086$	              &$2.207\pm0.0067$	  &$3.399\pm0.0036$	 &$3.735\pm0.0100$  \\                                           
 $\phi$			                & $1.1087\pm0.013$	                &$1.509\pm0.0176$	             &$1.832\pm0.0342$                   &$2.05\pm0.0220$	  &$3.240\pm0.02010$	  &$3.460\pm0.0480$ \\
 $\Lambda$				 &$1.297\pm0.0043$	              &$1.955\pm0.00943$	             &$2.487\pm0.021	$                      &$2.866\pm0.0380$	   &$3.807\pm0.0094	$         &$4.247\pm0.0313$  \\
 $\Xi^-$					& $1.097\pm0.013$              &	$1.614\pm0.0148$	             &$1.949\pm0.0320$	               &$2.170\pm0.0230$	     &$3.320\pm0.0340$	   &$3.509\pm0.0456$  \\
 $\Omega^-$				& $0.894\pm0.062$	           &$1.421\pm0.092$	                      &$1.877\pm0.279$	                &$1.958\pm0.0780$	     &3$.053\pm0.0340$	    &$3.241\pm0.0878$\\
  \hline
 \hline
\end{tabular}
\end{table*}
\end{center}
\begin{figure}[htbp]
		\centering
		\includegraphics[width=3.5in]{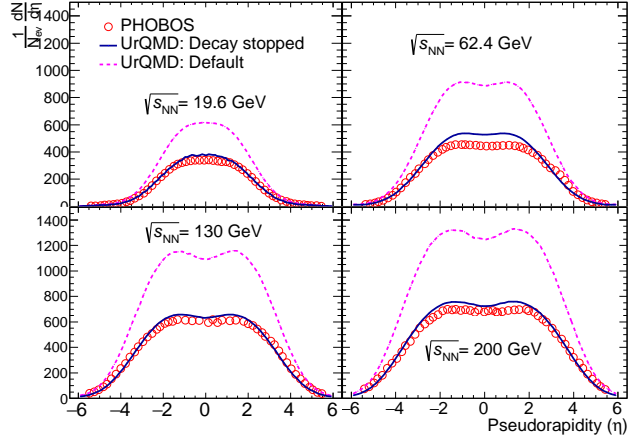}
		\caption{(Color online) Pseudorapidity distributions of all charged particles in Au-Au collisions at different RHIC energies $\sqrt{\it{s}_{NN}}= 19.6, 62.4, 130$, and $200$ GeV ($3\%$ central) and compared with existing experimental results of PHOBOS [\onlinecite{Alver}]. The red open circular symbols represent the experimental results of PHOBOS, whereas the magenta dotted and blue solid curves denote the UrQMD-3.4-generated data with default mode and switching off the weak decay channels.}
		\label{Fig:etaPHOBOS}	
\end{figure}

\begin{figure}[htbp]
		\centering
		\includegraphics[width=3.5in]{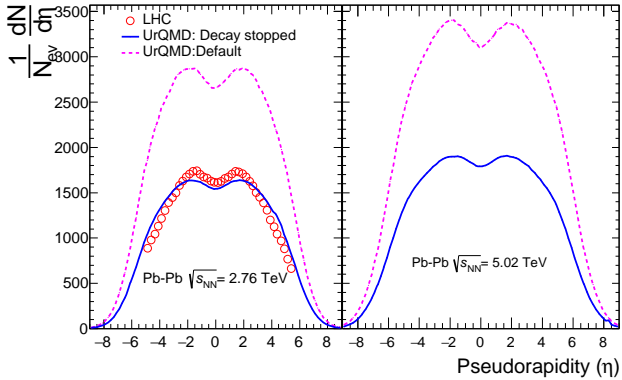}
		\caption{(Color online) Pseudorapidity distributions of all charged particles in Pb-Pb collisions at LHC energies,  $\sqrt{\it{s}_{NN}}= 2.76$ (left panel) and $5.02$ TeV (right panel) for $0-5\% $ centrality and compared with existing experimental results of ALICE at 2.76 TeV [\onlinecite{Abbas}]. The red open circular symbols represent the experimental results of ALICE, whereas the magenta dotted and blue solid curves denote the UrQMD-3.4-generated data with default mode and switching off the weak decay channels. }
		\label{Fig:etaLHC}	
\end{figure}

\begin{figure}[htbp]
		\centering
		\includegraphics[width=3.5in]{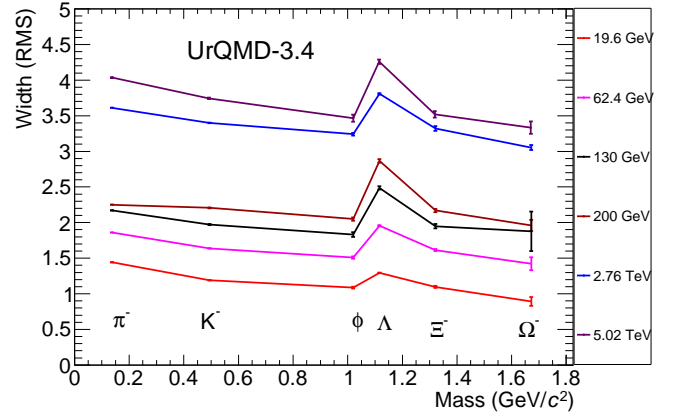}
		\caption{(Color online) Width of the rapidity distribution of various produced particles as a function of their masses for Pb-Pb collisions at RHIC and LHC energies using UrQMD-3.4-generated events. The solid lines with different colors correspond to various beam energies.}
		\label{Fig:widthRHICLHC}	
\end{figure} 

\begin{figure}[htbp]
		\centering
		\includegraphics[width=3.5in]{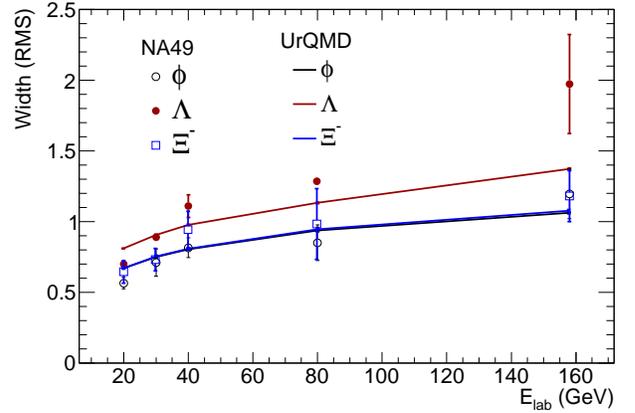}
		\caption{(Color online) Variation of the width of the rapidity distribution of various produced particles as a function of collision energy for Pb-Pb collisions at all the SPS energies for both NA49 and UrQMD-3.4-generated events. The solid lines and different marker symbols with different colors respectively denote the UrQMD-3.4-generated data and experimental results of various identified particles.}
		\label{Fig:widthSPSwithEnergy}	
\end{figure}
\begin{figure}[htbp]
		\centering
		\includegraphics[width=3.5in]{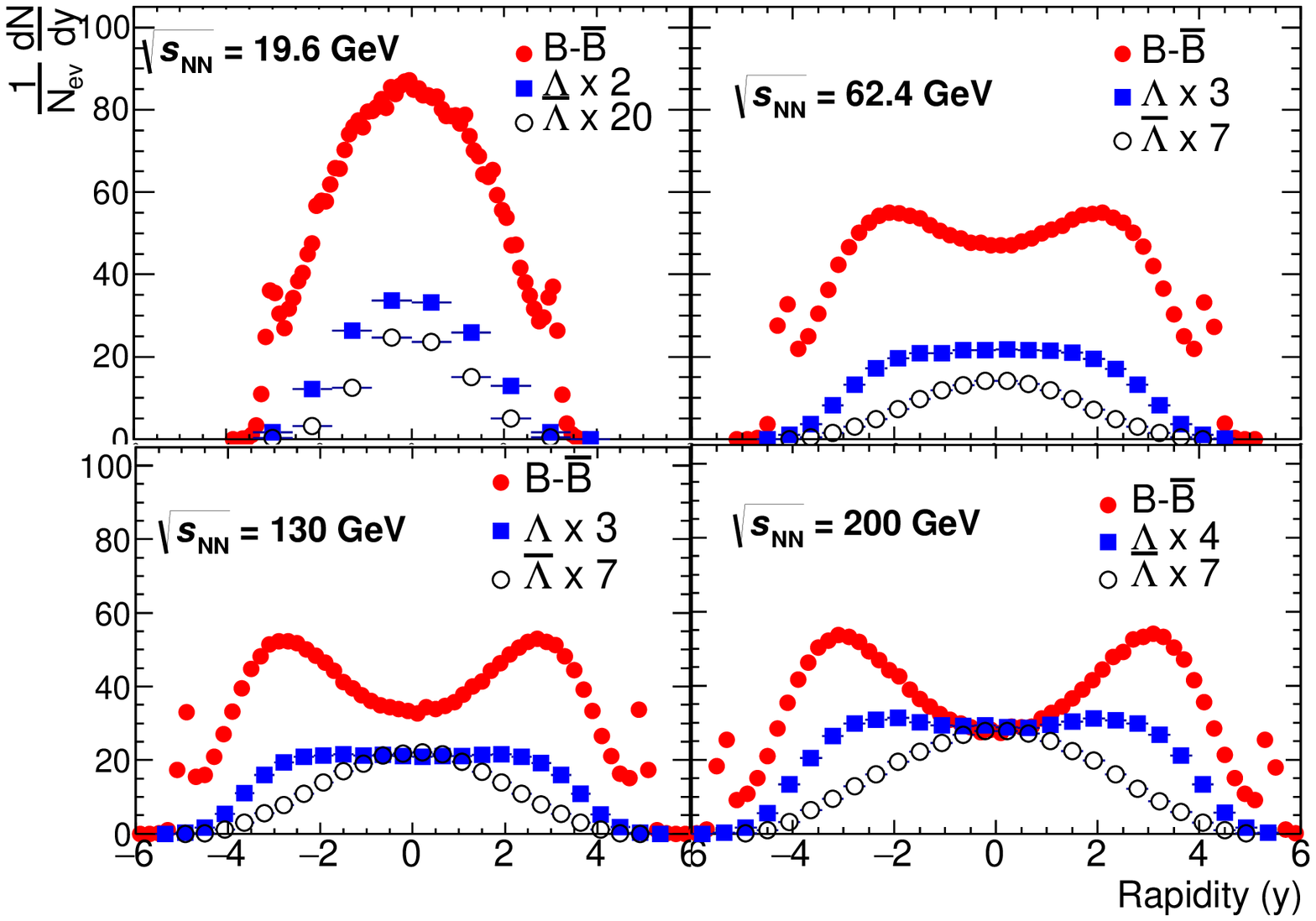}
		\caption{(Color online) Distributions of $B-\bar{B}$, $\Lambda$, and $\bar{\Lambda}$ over the rapidity space at RHIC energies using UrQMD-3.4-generated events. The filled circular symbols denote the $B-\bar{B}$ rapidity distributions, whereas the filled squares and open circular symbols respectively represent the rapidity distributions of $\Lambda$ and $\bar{\Lambda}$. }
		\label{Fig:netbaryonPhobos}	
\end{figure}

Because of the obvious reason of limited detector acceptance, no report on rapidity distribution of identified particles at RHIC and LHC energies could be found in the literature and hence the possible existence of the bumps in the experimental data of rapidity width of $\Lambda$ at these energies could not be ascertained. In the following, an attempt is made with MC data to examine if this behavior persists at RHIC and LHC energies too. It may be mentioned here that there are results of PHOBOS and ALICE experiments on pseudorapidity distributions of all charged particles at various RHIC and LHC energies. Both these experimental results on pseudorapidity distributions of unidentified charged particles excluded all contributions from the weak decays. To check how far the UrQMD-3.4 model is successful in describing the phase space evolution of particles produced in such heavy ion collisions at RHIC and LHC energies, two additional sets of central collision events have been generated for all the studied RHIC and LHC energies using UrQMD-3.4 event generator with default mode as well as by switching off all such weak decays. The model predicted pseudorapidity distributions of all charged particles in Au-Au collisions at $\sqrt{\it{s}_{NN}} = 19.6, 62.4, 130$, and $200$ GeV and in Pb-Pb collisions for $2.76 $ TeV and $5.02$ TeV are compared with the existing results of PHOBOS and ALICE experiments [\onlinecite{Mitrovski},\onlinecite{Abbas}] and are shown in Figs.~\ref{Fig:etaPHOBOS} and \ref{Fig:etaLHC} respectively.

Even though the default mode is found to overestimate the experimental values of pseudorapidity distribution significantly over the entire pseudorapidity range, the pseudorapidity spectra of generated events, excluding weak decays, are found to be in good agreement with the experimental results of both RHIC and LHC energies. The agreement between the UrQMD-3.4-model-generated data and experimental data is found to be much better than the earlier works of Refs.~[\onlinecite{Mitrovski}] and [\onlinecite{Abbas}]. The right panel of Fig.~\ref{Fig:etaLHC} shows the pseudorapidity spectra of all charged particles at 5.02 TeV for Pb-Pb collisions with model-generated data to date; no published result at this energy is available with experimental data thus far.

Using MC data with weak decay switched off, rapidity spectra of various identified particles are drawn for various RHIC and LHC energies and the widths of the rapidity distribution of each of the studied hadron are estimated as the r.m.s. values of the fitting function (\ref{eq:1}); the values are tabulated in the Table~\ref{tableRHICLHCWidth}.

The UrQMD-3.4 model predicted rapidity widths as functions of masses of the studied mesons and baryons in Au-Au and Pb-Pb collisions at RHIC and LHC energies are shown in Fig.~\ref{Fig:widthRHICLHC}. As seen at AGS and SPS energies, a clear jump in the width of the rapidity distribution of $\Lambda$ could be seen at all the RHIC and LHC energies. Such observation confirms that the enhanced width of rapidity distribution of $\Lambda$ is a characteristic feature of heavy ion collision data from AGS [\onlinecite{Deysir}] and SPS to RHIC and highest available LHC energies. For $\Xi^-$ ($dss$) containing only one leading quark, the rapidity width is seen to be more or less equal to the rapidity width of $\phi$ (Fig.~\ref{Fig:widthSPSwithEnergy}), compensating for the broadening due to net baryon density effect by the mass-dependent kinematic narrowing effect. However, the separate mass ordering for the studied mesons and baryons are seen to exist even at RHIC and LHC energies.
\begin{figure}[htbp]
		\centering
		\includegraphics[width=3.5in]{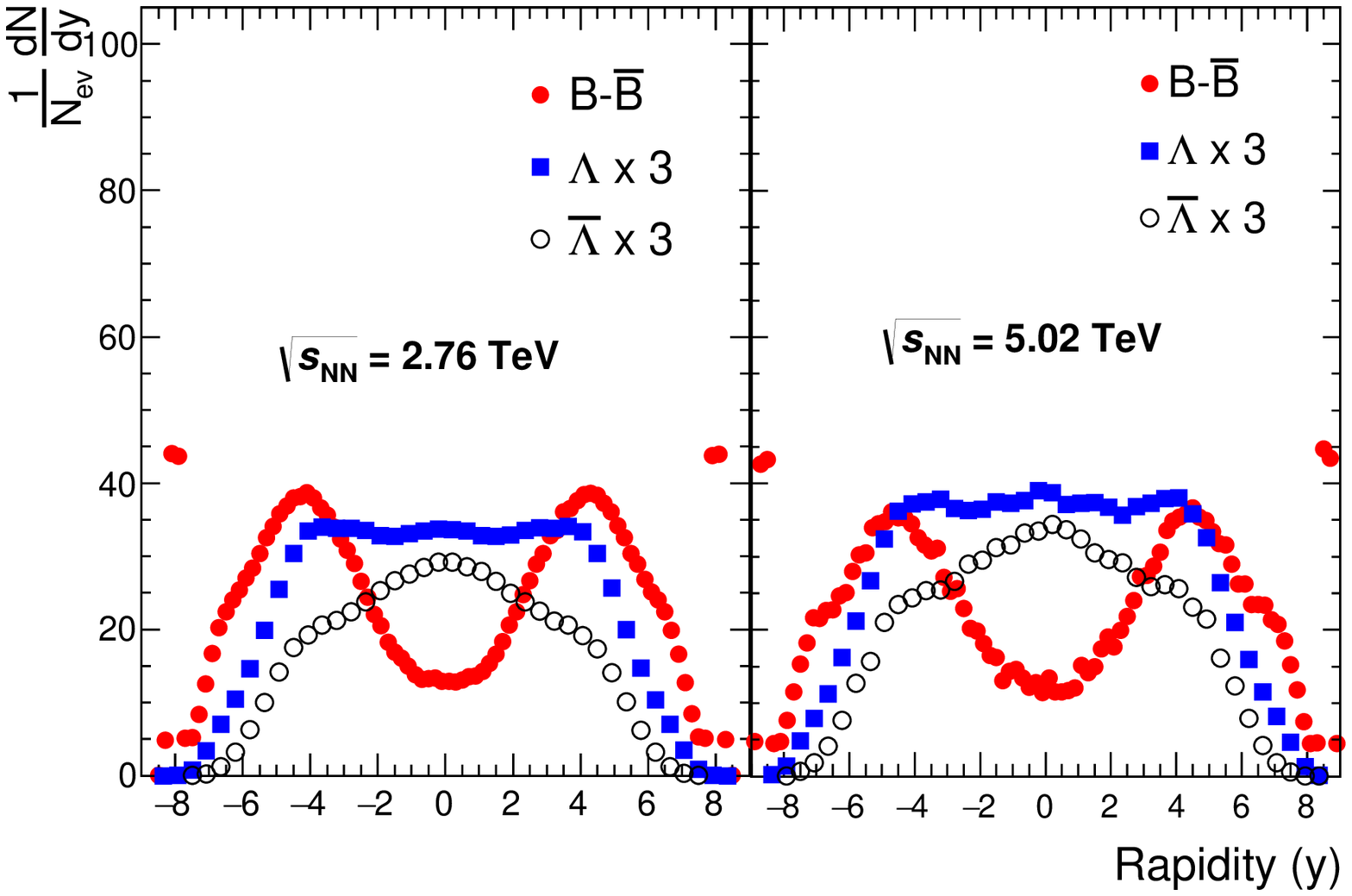}
		\caption{(Color online) Distributions of $B-\bar{B}$, $\Lambda$, and $\bar{\Lambda}$ over the rapidity space at $\sqrt{\it{s}_{NN}} = 2.76$ and $5.02$ TeV using UrQMD-3.4-generated events. The filled circular symbols denote the $B-\bar{B}$ rapidity distributions, whereas the filled squares and open circular symbols respectively represent the rapidity distributions of $\Lambda$ and $\bar{\Lambda}$. }
		\label{Fig:netbaryonLHC}	
\end{figure}

\section{\label{sec:level1}SUMMARY}
From the study on the width of the rapidity distribution of identified particles such as $\pi^-,  K^-, \phi, \Lambda, \Xi^-$, and $\Omega^-$  in Pb-Pb collisions at all the SPS energies using UrQMD-3.4-model-generated data and experimental data, it is readily seen that an increase in the width of the rapidity distribution of $\Lambda$  does exist at all the SPS energies from $20$ to $158\it{A}$ GeV. 

UrQMD-3.4-model-generated data, with weak decays stopped, could be seen to describe well the experimental pseudorapidity spectra both at RHIC and LHC energies. With generated data, the jump in the rapidity width of  $\Lambda$  could be seen at all the RHIC and LHC energies as well, thereby indicating its universal nature from the AGS and SPS to RHIC and highest LHC energies. To check if the rapidity distribution of $\Lambda$ (and $\bar{\Lambda}$) is still dependent on the net baryon density distribution even at RHIC and LHC energies, the distributions of $B-\bar{B}$, $\Lambda$, and $\bar{\Lambda}$ over the rapidity space at these energies are plotted in Figs.~\ref{Fig:netbaryonPhobos} and \ref{Fig:netbaryonLHC} respectively using UrQMD-3.4-generated events. It could be readily seen from these figures that just like at AGS and SPS energies [\onlinecite{Deysir}], the rapidity distribution of $\Lambda$ ($uds$), but not $\bar{\Lambda}$ ($\bar{u}\bar{d}\bar{s}$), takes the shape of net baryon density distribution even at RHIC and LHC energies. It is known that in heavy ion collision, a strange hadron like $\Lambda$ is produced either through associated strangeness production in $NN$ collisions of incoming and produced hadrons or through pair production [\onlinecite{SABAss},\onlinecite{DeySirNew}]. While the associated strangeness production is dominant at relatively lower incident beam energies, the pair production mechanism is considered dominant at higher incident beam energies. With the increase of transparency of collision, the net baryon number tends to populate the extreme rapidity regions, thereby increasing the population of baryons (like $\Lambda$) over antibaryons (like $\bar{\Lambda}$) at higher rapidity regions, which in turn increases the width of the rapidity distribution of particles [containing leading quark(s)] over antiparticles [containing produced quark(s) only]. Thus, such dependence of rapidity width of $\Lambda$ at RHIC and LHC energies, as seen from Figs.~\ref{Fig:netbaryonPhobos} and \ref{Fig:netbaryonLHC}, suggests that pair production may not be the only mechanism of particle production in heavy ion collisions even at the highest LHC energy. However, with MC data, over the entire studied energy range, no sign of violation of separate	 mass ordering of the width of rapidity distribution of mesons and baryons could be seen for the studied hadrons. 
\begin{acknowledgments}
The authors thankfully acknowledge the Department of Science and Technology (DST), Government of India, for providing funds through Project No. SR/MF/PS-01/2014-GU (C) to develop a high-performance computing cluster (HPCC) facility, which has been used to generated various Monte Carlo events for this work. One of the authors, Nur Hussain, would like to thank the DST, Government of India, for providing financial support in the form of a JRF scholarship. 
\end{acknowledgments}

\end{document}